\documentstyle[11pt,twoside,jltp]{article}
\title{Evolution of a Network of Vortex Loops in the Turbulent Superfluid
Helium; Derivation of the Vinen Equation}
\author{Sergey K.Nemirovskii
\address{Institute for Thermophysics, Prosp. Lavrentyeva,
1, Novosibirsk, 630090, Russia; E-mail: nemir@itp.nsc.ru}}
\runninghead{Sergey K.Nemirovskii}{Evolution of a Network of
Vortex Loops}
\begin{document}
\begin{abstract}
The evolution a network of vortex loops due to the fusion and breakdown in
the turbulent superfluid helium is studied. We perform investigation on the
base of the "rate equation" for the distribution function $n(l)$ of number
of loops in space of their length $l$. There are two mechanisms for change
of quantity $n(l)$. Firstly, the function changes due to deterministic
process of mutual friction, when the length grows or decreases depending on
orientation. Secondly, the change of $n(l)$ occurs due to random events when
the loop crosses itself breaking down into two daughter or two loops collide
merging into one larger loop. Accordingly the "rate equation" includes the
"collision" term collecting random processes of fusion and breakdown and the
deterministic term. Assuming, further, that processes of random colliding
are fastest we are in position to study more slow processes related to
deterministic term. In this way we study the evolution of full length of
vortex loops per unit volume-so called vortex line density ${\cal L}(t)$. It
is shown this evolution to obey the famous Vinen equation. In conclusion we
discuss properties of the Vinen equation from the point of view of the
developed approach.
\end{abstract}
\maketitle
\section{INTRODUCTION AND SCIENTIFIC BACKGROUND}
In spite of very long history \ the theory superfluid turbulence is very far
from the more or less completeness (\cite{Feynman}\cite{Vinen},\cite{NF},).
The reason to this is an incredible complexity of this problem. Indeed we
have to deal with a set of objects (vortex loops) which do not have a fixed
number of elements, they can split and merge in processes of reconnection.
Thus, some analog of the secondary quantization method is required with the
difference that the objects (vortex loops) themselves possess an infinite
number of degrees of freedom with very involved dynamics. Clearly this
problem can hardly be resolved in the nearest future. As for direct
numerical simulations, which remain the main source of information about
this process \cite{Schwarz88}\cite{Barenghi},\cite{Aarts}, \cite{Tsubota00},
there are also many problems. It is clear that any progress can be achieved
only if one can essentially reduce the number of degree of freedom of each
of the loops. One way elaborated in context of lambda-transition (\cite%
{Williams}) is to imagine vortex loops as a set of rings of different sizes
and to take their radius as the only degrees of freedom. In paper of author
(see \cite{Nemirovskii_97_1}) there is offered another way - to think of
vortex loops as randomly walking chains.The physical ground for this
supposition is the following one. During evolution the loops undergo huge
amount of collisions with other loops (or with itself) \ with consequent
reconnection. As a result each of the loops consists of many uncorrelated
parts of mean size $\xi _{0}$ so called elementary step.A bit more precisely
an average loop consists of smoothly (but randomly) connected arcs of mean
radius $\xi _{0}.$ This fact allows to consider vortex loop as a random
walk, and to describe its statistics with help of some generalized Wiener
distribution. Parameters of this probability distribution functional can be
regarded as degrees of freedom of vortex loop. This conception allowed to
fulfil a series of studies on the vortex tangle properties, e. g. to
calculate its momenta, energy \ and spectrum of energy. In the present work
we demonstrate how this point conception allows to derive analytically the
famous Vinen equation for evolution of the vortex line density ${\cal L}(t).$
\section{KINETIC EQUATION}
Following the work \cite{Copeland98} we introduce the distribution function $%
n(l,t)$ of the density of a loop in the \textquotedblright
space\textquotedblright\ of their lengths. It is defined as the number of
loops (per unit volume) with lengths lying between $l$ and $l+dl$. \ There
are two mechanisms for change of $n(l,t).$ One of them is the deterministic
process. In superfluid helium quantity $n(l,t)$ changes due to interaction
of vortices with the normal component. This processes is more or less clear
and can be modeled by relatively simple equations of motion. Other reasons
for change of quantity $n(l,t)$ are the random reconnection processes. We
discriminate two types of the reconnection processes, namely the fusion of
two loops into the larger single loop and the breakdown of single loop into
two daughter loops. The kinetics of the vortex tangle is affected by the
intensity of the introduced processes. The intensity of the first process is
characterized by the rate $A(l_{1},l_{2},l)$ (number of events per unit time
and per unit volume) of collision of two loops with lengths $l_{1}$ and $%
l_{2}$ and forming the loop of length $\ l=l_{1}+l_{2}$. The intensity of
the second process is characterized by the rate of self-intersection $%
B(l,l_{1},l_{2})$ of the loop with length $l$ into two daughter loops with
lengths $\ l_{1}$ and $l_{2}$. In view of what has been exposed above we can
directly write out the master \textquotedblright kinetic\textquotedblright\
equation for rate of change the function $n(l,t)$.
\begin{eqnarray}
\frac{\partial n(l,t)}{\partial t}+\frac{\partial n(l,t)}{\partial l}\frac{%
\partial l}{\partial t}=\int \int A(l_{1},l_{2},l)n(l_{1})n(l_{2})\delta
(l-l_{1}-l_{2})dl_{1}dl_{2}\;\;\;\;\;\;\;\;\;\;\;\ \ \ \   \nonumber \\
-\int \int A(l_{1},l,l_{2})n(l)n(l_{1})\delta
(l_{2}-l_{1}-l)dl_{1}dl_{2}\;\;\;\;\;\;\;\;\;\;\ \ \ \ \ \ \ \ \ \;\ \ \ \ \
\ \;\;  \nonumber \\
-\int \int A(l_{2},l,l_{1})n(l)n(l_{2})\delta
(l_{1}-l_{2}-l)dl_{1}dl_{2}\;\;\;\;\;\;\;\;\;\;\;\;\ \ \ \ \ \ \ \ \ \ \ \ \
\ \   \nonumber \\
-\int \int B(l_{1},l_{2},l)n(l)\delta
(l-l_{1}-l_{2})dl_{1}dl_{2}\;\;\;\;\;\;\;\;\;\ \ \ \;\;\;\ \;\;\;\;\;\;\;\;\
\ \ \ \ \ \ \   \label{Master} \\
+\int \int B(l,l_{2},l_{1})n(l_{1})\delta
(l_{1}-l-l_{2})dl_{1}dl_{2}\;\;\;\;\;\;\;\ \;\;\;\;\;\;\;\;\;\;\ \ \ \ \ \ \
\ \ \ \;  \nonumber \\
+\int \int B(l,l_{1},l_{2})n(l_{1})\delta
(l_{2}-l-l_{1})dl_{1}dl_{2}\;.\;\;\;\
\;\;\;\;\;\;\;\;\;\;\;\;\;\;\;\;\;\;\;\;\ \ \ \ \ \ \ \ \ \ \ \ \ \
\;\;\;\;\;\;  \nonumber
\end{eqnarray}
Coefficients $A$ and $B$ are calculated in \cite{Nemir_05_RR}
\[
A(l_{1},l_{2},l)=b_{m}V_{l}l_{1}l_{2},\;\ \ B(l_{1},l_{2},l)=b_{s}V_{l}l(\xi
_{0}l_{1})^{-3/2}.
\]%
Here $b_{m}$ and $b_{s}$ are some constants of order of unity. In paper \cite%
{Nemir_05_RR} on vortex loops in superfluid turbulent HeII there was offered
$b_{m}\approx 1/3,$ $b_{s}\approx 0.0164772.$ Quantity $V_{l}$ is the
characteristic velocity of lines by order magnitude equal $\kappa /\xi _{0}$
($\kappa $ is the quantum of circulation). The Brownian random walk approach
fails for scales near $\xi _{0},$ therefore usually this value appears as a
low cut-off. In paper \cite{NemirPRL06} \ there was shown that \ the
collision term alone leads to steady solution expressed by formula
\begin{equation}
n(l)=Cl^{-5/2}.  \label{n(l)}
\end{equation}%
Solution (\ref{n(l)}) is nonequilibrium solution with the $l$ -independent
flux $P$ of the length density $L=l\ n(l)$ , which is the length accumulated
in loops of length lying between $l$ and $l+dl$. Term
"flux" here means just the redistribution of length among
the loops due to reconnections. Quantity $P$ is equal
\begin{equation}
P=6.\,\allowbreak 277\,5C^{2}b_{m}\kappa /\xi _{0}-2.\,\allowbreak
772\,5Cb_{s}\kappa \xi _{0}{}^{-5/2}.  \label{flux}
\end{equation}%
The master equation (\ref{Master}) is the base for study of
superfluid turbulence in the frame of conception of randomly walking chains.
The aim of our present paper is to demonstrate that the vortex line density,
total length per unit volume defined as ${\cal L}(t)=\int n(l)ldl$ \ evolves
in accordance with the famous Vinen equation\cite{Vinen}.
\section{VINEN EQUATION}
To show it let us multiply the kinetic equation (\ref{Master}) by$\
l$ and integrate over all sizes
\begin{equation}
\frac{d{\cal L}(t)}{dt}=\int \frac{\partial n(l,t)}{\partial t}ldl=-\int
\frac{\partial n(l,t)}{\partial l}\frac{\partial l}{\partial t}%
ldl-P_{net}.\;\;  \label{VE_0}
\end{equation}%
Here the quantity $P_{net}$ is the net \textquotedblright
flux\textquotedblright\ of length in $l-$space, which is absolute value of
flux $P$ expressed by relation (\ref{flux}). This rule is imposed because
flux \ $P$ always carries away the vortex line density ${\cal L}$ \ from the
system, and different signs refers to direction of the cascade.
First we treat the deterministic term in equation (\ref{VE_0}). Due to
enormous number of reconnection it is quite natural to suppose that
"equilibrium" solution (\ref{n(l)}) is reached much faster than the slow
deterministic change of function $n(l).$ Then we can use this solution to
evaluate deterministic term. We will calculate the rate of change of length
of each loop on the base of the motion equation of \ the line in local
approach (see e.g. \cite{Schwarz88})%
\begin{equation}
{\bf v}_{l}=\beta {\bf s}^{\prime }\times {\bf s}^{\prime \prime }+\alpha
{\bf s}^{\prime }\times ({\bf V}_{ns}-\beta {\bf s}^{\prime }\times {\bf s}%
^{\prime \prime })  \label{line_velocity}
\end{equation}%
Here ${\bf s}^{\prime }$ and ${\bf s}^{\prime \prime }$ are the first and
second derivatives from position of line ${\bf s}(\xi )$ with respect to
label variable $\xi $, which coinsides here with the arclength. to
calculate, ${\bf v}_{l}(\xi )$ is velocity of the line element.
To calculate $\partial l/\partial t,$ we use the relation for the rate of
change of length $\partial \delta l/\partial t$ \ for some arbitrary element
with length $\delta l~$(see.g. \cite{Schwarz88}). Assuming for a while that
the label variable $\xi $ is not exactly the arclength, we have $\delta
l=\left\vert {\bf s}^{\prime }\right\vert $ $\delta \xi .~\ $Then\
the \ following chain of relations takes place.%
\[
\frac{\partial \delta l}{\partial t}=\frac{\partial \left\vert {\bf s}%
^{\prime }\right\vert \delta \xi }{\partial t}=\frac{\left\vert {\bf s}%
^{\prime }\right\vert }{\left\vert {\bf s}^{\prime }\right\vert }\frac{%
\partial \left\vert {\bf s}^{\prime }\right\vert \delta \xi }{\partial t}=%
\frac{{\bf s}^{\prime }}{\left\vert {\bf s}^{\prime }\right\vert }\frac{%
\partial {\bf s}^{\prime }\delta \xi }{\partial t}={\bf s}^{\prime }{\bf v}%
_{l}^{\prime }\delta \xi
\]%
On the last stage we return to $\left\vert {\bf s}^{\prime }\right\vert =1.$%
Diffirentiating (\ref{line_velocity}) and multiplying by ${\bf s}^{\prime }$
we have after little algebra
\begin{equation}
\frac{\partial\delta l}{\partial t}=(\alpha({\bf s}^{\prime}\times {\bf s}%
^{\prime\prime}){\bf V}_{ns}-\alpha\beta({\bf s}^{\prime }\times{\bf s}%
^{\prime\prime})^{2})\delta\xi  \label{ddl/dt}
\end{equation}
The next step is to average expression (\ref{ddl/dt}) over all possible
configurations of vortex loops. We do it with use of the Gaussian model of
the vortex tangle elaborated earlier by author \cite{Nemirovskii_97_1}. In
accordance with this model
\begin{equation}
\langle{\bf s}^{\prime}\times{\bf s}^{\prime\prime}\rangle\;=\frac {I_{l}}{%
\sqrt{2}c_{2}\xi_{0}}\frac{{\bf V}_{ns}}{\left\vert {\bf V}_{ns}\right\vert }%
,\ \ \ \ \langle({\bf s}^{\prime}\times{\bf s}^{\prime\prime})^{2}\rangle\;=%
\langle({\bf s}^{\prime\prime})^{2}\rangle=\frac{1}{2\xi_{0}^{2}}.\ \ \ \ \
\label{s's''}
\end{equation}
Quanity $c_{2}$ is the structure constant introduced by Schwarz \ \cite%
{Schwarz88}). \ It follows from (\ref{ddl/dt}) and from (\ref{s's''}) that $%
\partial \delta l/\partial t\propto \l $. Substituting (\ref{s's''}) into
(averaged equation (\ref{ddl/dt})) and then into (\ref{VE_0}) and
integrating by part we get the contribution into $d{\cal L}(t)/dt\ $from the
deterministic term.
\begin{equation}
(\alpha \frac{I_{l}\left\vert {\bf V}_{ns}\right\vert }{\sqrt{2}c_{2}\xi _{0}%
}-\alpha \beta \frac{1}{2\xi _{0}^{2}})\int \frac{\partial n(l,t)}{\partial l%
}l^{2}dl=-\alpha \frac{2I_{l}\left\vert {\bf V}_{ns}\right\vert }{\sqrt{2}%
c_{2}\xi _{0}}{\cal L+}\frac{\alpha \beta }{\xi _{0}^{2}}{\cal L}.
\label{det_rate_L}
\end{equation}
Now we have to treat the "flux" term in equation (\ref{VE_0}).
Unlike paper \cite{NemirPRL06} where the constant $C$ \ in relation (\ref%
{n(l)}) was obtained from condition $P=0$, here we obtain $C$ from the
normalization condition ${\cal L}(t)=\int n(l)ldl=C\int l^{-3/2}dl=2C/\xi
_{0}$ (We recall that $\xi _{0}$ is a low cut-off of the wholle approach and
integral diverges on low limit ). Furthermore we consider consequently the
collision and reconnection events\ to put the system in equilibrium (with
respect to solution (\ref{n(l)})) state much faster than the slow
deterministic processes. This implies that parameters $\xi _{0}$ and $I_{l}$%
, the so called structure constants of the vortex tangle, have a time to
adjust to their equilibrium values. This assumption is widely adopted and it
was confirmed in numerical simulations in \cite{S_Rozen}. In particular the
mean radius of curvature (playing a role of elementary step in our approach)
is related to the vortex line density as $\xi _{0}^{2}=1/2\;c_{2}^{2}\;{\cal %
L}$ (see \cite{Schwarz88}\cite{Nemirovskii_97_1},). \ By use of the said
above we rewrite expression for flux (\ref{flux}) in form $P=C_{F}\kappa
{\cal L}^{2},$ where the temperature constant $C_{F}$ is
\begin{equation}
C_{F}=(1.\,\allowbreak 569\,4b_{m}-2.\,\allowbreak
772\,6c_{2}^{2}\allowbreak b_{s}).  \label{CF}
\end{equation}%
We named this constant in honor of Feynman who was the first person to
discuss evolution of vortex line density due to the reconnection processes.
We would like to recall that he supposed decay of the vortex tangle due to
the cascade like breakdown of vortex loops with further disappearance of
them on very small scales. Relation (\ref{CF}) shows that there is possible
the inverse cascade, which corresponds to the cascade-like fusion of \
vortex loops. Unfortunately our approach has too approximate character to do
any strong quantitative conclusion. It is clear, however, that for low
temperatures, where the vortex tangle is more kinky, correspondingly $c_{2}$
is large, the quantity $C_{F}$ \ is negative. This corresponds to the direct
cascade in region of very small loops. On the contrary for \ high
temperature lines are smoother, $c_{2}$ is small, and $C_{F}$ \ is positive,
which implies that there is inverse cascade with formation of large loops.If
we for instance adopt values for $b_{m}$ and $b_{s}$ and use for $c_{2}^{2}$
values offered by Schwarz (see \cite{Schwarz88}), then we get $C_{F}\
\approx -0.01$ for the temperature $1.07$ K and $\ C_{F}\approx 0.44$ K for
the temperature $2.01$ K.
Collecting contribution into $d{\cal L}(t)/dt$ (\ref{VE_0}) from both the
deterministic and collision processes, and taking into account that $%
P_{net}=\left\vert C_{F}\right\vert \kappa \allowbreak {\cal L}^{2}$ we
finally have%
\begin{equation}
\frac{d{\cal L}(t)}{dt}=\frac{5}{2}\alpha I_{l}\left\vert V_{ns}\right\vert
{\cal L}^{\frac{3}{2}}-\frac{5}{2}\alpha \beta c_{2}^{2}{\cal L}%
^{2}-\left\vert C_{F}\right\vert \kappa \allowbreak {\cal L}^{2}.
\label{VE_2}
\end{equation}%
Thus, starting with kinetics of a network of vortex loops, we get the famous
Vinen equation\cite{Vinen}. Let us discuss meaning of various terms entering
this equation. The first, generating term in the rhs of the Vinen equation
describes the grows of the vortex tangle due to mutual friction. The second
term also connected to mutual friction, however this term is responsible for
decrease of the vortex line density. This point of view coinsides with ideas
by Schwarz \cite{Schwarz88} who obtained the deterministic contribution into
$d{\cal L}(t)/dt$ using a bit different approach.
The third term in the RHS of (\ref{VE_2}) is related to random collisions of
vortex loops. It describes decrease of the vortex line density due to the
flux of length carrying away the length from the system. Depending on
interplay between coefficients $b_{m}$ and $b_{s}$ and parameters $c_{2}$ \
the flux can be either positive or negative. Negative flux appears when
break down of loops prevails and cascade-like process of generation of
smaller and smaller loops forms. There exists a number of mechanisms of
disappearance of rings on very small scales. It can be e.g. acoustic
radiation, collapse of lines, Kelvin waves etc. Thus, in this case the
situation is fully coincides with the scenario proposed by Feynman\cite%
{Feynman}.
The case when the flux is positive is less clear. Positive flux implies the
cascade-like process of generation of larger and larger loops. Unlike
previous case of negative flux, there is no apparent mechanism for
disappearance of very large loops. There are possible the following
possibilities. The first one is that very large loops, whose $3D$ size is
comparable with size of the volume (note that length of these loop is much
larger because of the random walk structure) can annihilate fully or partly
on boundaries. Then the picture reminds the Feynman scenario but with the
inverse flux. The second possibility is that parts of large loops is pinned
on the walls and the whole approach representing the line to have the random
walk structure fails. Finally it should not be ruled out, that a state with
lines pinned on the walls and stretching from wall to wall is just a
degenerate state of the vortex tangle. Some of numerical investigators\cite%
{Schwarz88}\cite{Aarts}, report on this situation.
\section*{ACKNOWLEDGMENTS}
This work was partially supported by grant N 05-08-01375 of the RFBR and
grant of the President of the Russian Federation on the state support of
leading scientific schools NSH-6749.2006.8.

\end{document}